\documentclass[
reprint,
showpacs,
 amsmath,amssymb,
 aps,nobibnotes, 
prl,
]{revtex4-1}

\usepackage{graphicx}
\usepackage{dcolumn}
\usepackage{bm}
\usepackage{hyperref}

\hypersetup{colorlinks,
	linkcolor=blue,
	filecolor=blue,
	urlcolor=blue,
	citecolor=blue
}

\usepackage[caption=false]{subfig}
\usepackage{xcolor}

\newcommand{\bra}[1]{\ensuremath{\langle #1 \rvert}}

\newcommand{\bvec}[1]{\ensuremath{\boldsymbol{#1}}}
\newcommand{\disav}[1]{\ensuremath{\overline{#1}}}
\newcommand{\eg}{{\it e.g.\ }}
\newcommand{\Emax}{E_{\mathrm{max}}}
\newcommand{\ie}{{\it i.e.\ }}
\newcommand{\ket}[1]{\ensuremath{  \lvert  #1  \rangle}}
\newcommand{\Lgp}{\mathcal{L}_{\mathrm{GP}}}
\newcommand{\vecr}{\bvec{r}}
\newcommand{\vecrp}{\bvec{r^\prime}}

\begin{document}

\title{Superfluid-insulator transition of two-dimensional disordered Bose gases}

\author{Joseph Saliba, Pierre Lugan, and Vincenzo Savona}

\affiliation{Institute of Theoretical Physics, Ecole Polytechnique F\'ed\'erale de Lausanne EPFL, CH-1015 Lausanne, Switzerland.}

\date{\today}


\begin{abstract}
We study the two-dimensional weakly repulsive Bose gas at zero temperature in the presence of correlated disorder. Using large-scale simulations, we show that the low-energy Bogoliubov cumulative density of states remains quadratic up to a critical disorder strength, beyond which a power law with disorder-dependent exponent $\beta<2$ sets in. We associate this threshold behavior with the transition from superfluid to Bose glass, and compare the resulting mean-field phase diagram with scaling laws and the Thomas-Fermi percolation threshold of the mean-field density profile.
\end{abstract}

\pacs{67.85.--d, 05.30.Jp, 03.75.Hh, 64.70.Tg}


\maketitle


Disorder can affect the properties of condensed-matter systems up to the point of completely suppressing transport, thereby driving metals~\cite{lee_disordered_1985}, superconductors~\cite{beloborodov_granular_2007} and superfluids~\cite{weichman_dirty_2008} into insulating phases. While the Anderson transition of single particles is now relatively well understood~\cite{evers_anderson_2008}, the combination of disorder and interactions still poses a number of important challenges~\cite{imada_metal-insulator_1998}. In systems of repulsive bosons with potential disorder, the low-temperature superfluid (SF) phase competes with a compressible insulator called Bose glass (BG) that prevails except for a regime of weak disorder and intermediate interaction strength~\cite{giamarchi_localization_1987, fisher_boson_1989}. The mechanisms of the SF-BG transition, which are relevant for ${}^4$He in porous media~\cite{crowell_superfluid-insulator_1995}, Josephson-junction arrays~\cite{vanderzant_quantum_1996}, arguably also driven-dissipative polariton fluids~\cite{manni_polariton_2011} and superconducting films~\cite{goldman_two-dimensional_1995}, have recently attracted renewed interest due to experiments with quantum magnets~\cite{zheludev_dirty-boson_2013} and ultracold gases~\cite{fallani_ultracold_2007, deissler_delocalization_2010, pasienski_disordered_2010, krinner_superfluidity_2013}.  Through unprecedented control over interactions and disorder statistics~\cite{sanchez-palencia_disordered_2010}, the latter allow quantitative comparisons with theory~\cite{jendrzejewski_three-dimensional_2012}.

On the theory side, progress has been made in understanding the features of the SF-BG transition in one dimension (1D), such as the detailed shape of the phase diagram~\cite{lugan_ultracold_2007, falco_weakly_2009, fontanesi_mean-field_2010, vosk_superfluid-insulator_2012, prokofev_comment_1998, rapsch_density_1999, pollet_absence_2009}, the critical regime~\cite{altman_insulating_2008, hrahsheh_disordered_2012, ristivojevic_phase_2012, pielawa_numerical_2013, pollet_asymptotically_2014}, the connection to finite temperature~\cite{aleiner_finite-temperature_2010}, and dynamical properties related to elementary excitations~\cite{gurarie_excitations_2008, fontanesi_superfluid_2009, fontanesi_mean-field_2010}. Yet, much less is known about two dimensions (2D), including the weakly interacting regime that is the focus of current experiments with ultracold Bose gases~\cite{beeler_disorder-driven_2012, allard_effect_2012}. Renormalization-group (RG) approaches are complicated by the lack of an equivalent of Luttinger-liquid theory~\cite{giamarchi_localization_1987} and by higher connectivity~\cite{iyer_mott_2012}, which reduces the impact of isolated weak links~\cite{fontanesi_mean-field_2010, altman_insulating_2008, pollet_asymptotically_2014, kane_transmission_1992}. The role of percolation, in particular, is supported by numerical RG and Gutzwiller studies~\cite{iyer_mott_2012, niederle_superfluid_2013}. Monte Carlo calculations have also successfully characterized the quantum phase diagram at strong interactions but appear to be challenged by the large system sizes required in the opposite limit~\cite{krauth_superfluid-insulator_1991, soyler_phase_2011}. Other recent studies addressed nonzero temperatures~\cite{carleo_universal_2013} and the special case of infinitely repulsive lattice bosons~\cite{zuniga_bose-glass_2013}. As for the weakly interacting regime, scaling laws have been derived on the basis of purely dimensional, mean-field considerations~\cite{falco_weakly_2009, fontanesi_mean-field_2010}, but further quantitative predictions and actual signatures of the $T=0$ transition are missing in that context, with the exception of the superfluid fractions analyzed in Ref.~\cite{astrakharchik_phase_2013}.

In this work, we examine the ground state of a disordered weakly interacting 2D Bose gas, and account for quantum fluctuations within Bogoliubov theory. Our analysis of the SF-BG transition involves Landau's criterion for superfluidity~\cite{landau_lifshitz_vol9}, which in the disorder-free case relates the stability of the SF phase against the creation of excitations, to their linear dispersion relation $E(k)$ at low energy, \ie to an $E^2$ energy dependence of their cumulative density of states (CDOS) in 2D. While momentum is not conserved in the disordered case, the energy dependence of the disorder-averaged CDOS still allows an inspection of Landau's criterion. Our numerical results reveal an abrupt change in the Bogoliubov CDOS that marks the suppression of phononlike excitations and the transition to the Bose glass. This signature is used to delineate the SF-BG phase diagram as a function of disorder amplitude and spatial correlation vs interaction strength, and to assess the predictions of a classical percolation analysis.


The Bose gas is described by the continuum many-body Hamiltonian
\begin{equation}
\hat H= \int d\vecr \left[ \hat \Psi^\dagger (\vecr) \hat H_0 \hat \Psi (\vecr)+ \frac{g}{2} \hat\Psi^\dagger (\vecr) \hat\Psi^\dagger (\vecr) \hat\Psi (\vecr) \hat\Psi (\vecr)\right],
\end{equation}
where $\hat \Psi(\vecr)$ is the bosonic field operator, $g>0$ is the coupling constant of a repulsive contact interaction, $\hat H_0 = -\frac{\hbar^2}{2m}\nabla_{\vecr}^2 + V(\vecr)$ is the noninteracting Hamiltonian, and $V(\vecr)$ is a random potential with configuration average $\disav{V(\vecr)}=0$. Although the approach below is general, we assume that $V$ is Gauss distributed and Gauss correlated, with $\disav{V(\vecr) V(\vecrp)} = \Delta^2 e^{-(\vecr-\vecrp)^2/2\eta^2}$, and we introduce the energy scale $E_c=\hbar^2/(2m\eta^2)$ associated with the correlation length $\eta$. For weak interactions, the Bose gas is accurately described by Bogoliubov theory~\cite{bogoliubov_1947}. In the density-phase formulation of the latter~\cite{mora_extension_2003}, the field operator is written as $\hat{\Psi}(\vecr)=e^{i\hat{\theta}(\vecr)}\sqrt{\rho_0(\vecr)+\delta\hat\rho(\vecr)}$, and $\hat{H}$ is expanded to second order in the quantum fluctuations $\hat{\theta}$ and $\delta\hat{\rho}$. The ground-state mean-field density $\rho_0(\vecr)$ obeys the Gross-Pitaevskii equation
\begin{equation}\label{eq:GPE}
[\hat{H}_0 + g \rho_0(\vecr)] \sqrt{\rho_0(\vecr)} =\mu \sqrt{\rho_0(\vecr)},
\end{equation}
where $\mu$ is the chemical potential. The quadratic Bogoliubov Hamiltonian is ``diagonalized''  by a canonical transformation to the bosonic quasiparticle operators $\hat{b}_j=\int d\vecr[u_j^*(\vecr)\delta\hat{\Psi}(\vecr)-v_j^*(\vecr)\delta\hat{\Psi}^\dagger(\vecr)]$, with $\delta\hat{\Psi}(\vecr)=i\sqrt{\rho_0(\vecr)}\hat{\theta}(\vecr)+\delta\hat{\rho}(\vecr)/[2\sqrt{\rho_0(\vecr)}]$, while $u_j$ and $v_j$ are given by the positive-energy ($E_j>0$) solutions of the Bogoliubov--de Gennes equation (BdGE)
\begin{align}\label{eq:BdGE}
\Lgp \begin{pmatrix}u_j(\vecr)\\v_j(\vecr)\end{pmatrix}
&=E_j\begin{pmatrix}u_j(\vecr)\\v_j(\vecr)\end{pmatrix}
\end{align}
with
\begin{align}\label{eq:LGP}
\Lgp&=\begin{pmatrix}
\hat{H}_0 +2g\rho_0(\vecr) -\mu & g\rho_0(\vecr)\\
-g\rho_0(\vecr) & -\hat{H}_0 -2g\rho_0(\vecr) +\mu
\end{pmatrix}.
\end{align}

Equations (\ref{eq:GPE})--(\ref{eq:LGP}) give access to the excitation spectrum, the correlation functions, and the thermodynamic properties of the weakly interacting Bose gas at low temperatures~\cite{mora_extension_2003, pitaevskii_book_2003}. For the $d$-dimensional Bose gas at $T=0$, the Bogoliubov expansion is valid wherever $\rho_0(\vecr)\xi^d\gg1$ \cite{castin_low-temperature_1998, mora_extension_2003}. Here $\xi=\hbar/\sqrt{mU}$ is the healing length associated with the interaction energy $U=g\disav{\rho_0}$. Hence, even in the strongly disordered case, the regions of space where Bogoliubov theory breaks down become asymptotically small in the limit $\disav{\rho_0}\to\infty$ ($g\to 0$) at constant $U$. Remarkably, the derivation of the BdGE in the density-phase picture does not rely on the existence of a condensate with well-defined phase, but rather on the smallness of phase gradients and relative density fluctuations, as ensured by the sole small parameter $1/(\disav{\rho_0}\xi^d)$. The density-phase formulation has thus successfully been used to describe quasicondensate phases~\cite{mora_extension_2003}, as well as properties of the insulating phase across the SF-BG transition~\cite{fontanesi_superfluid_2009}. In the latter setting, the Bogoliubov approximation neglects (subleading) corrections in inverse powers of the density that become relevant in a many-body description of the critical regime~\cite{vosk_superfluid-insulator_2012, pollet_asymptotically_2014, iyer_mott_2012}. Yet, for typical experiments in the weakly interacting regime~\cite{deissler_correlation_2011, allard_effect_2012}, where criterion $\disav{\rho_0}\xi^d\gg1$ is met by one or two orders of magnitude, many-body corrections to the phase boundary are expected to remain small. Interestingly, the asymptotic proportionality of the critical Luttinger parameter $K_c$ to $\disav{\rho_0}$ in the mean-field limit ($\disav{\rho_0}\to\infty$ at fixed $U$), as inferred from Refs.~\cite{fontanesi_superfluid_2009, fontanesi_mean-field_2010, saliba_superfluid-insulator_2013}, also appears to be compatible with the nonuniversal behavior $K_c\neq3/2$ recently put forward for the weakly interacting and strongly disordered regimes in 1D~\cite{altman_insulating_2008, pollet_asymptotically_2014}. Besides, the existence of a true critical behavior at the mean-field level, as suggested for 1D~\cite{pollet_classical-field_2013}, remains an open question that calls for an analysis within Bogoliubov theory in dimension $d>1$.

In the present work, we analyze the CDOS of Bogoliubov excitations, defined as
\begin{align}
\mathcal{N}_{\mathcal{V}}(E)&=\frac{1}{\mathcal{V}}\int_{0^+}^E dE' \sum_j \delta(E'-E_j)
\end{align}
for a system of volume $\mathcal{V}$ and a given potential configuration, with $E_j$ the corresponding eigenvalues of the BdGE~(\ref{eq:BdGE}). In the thermodynamic limit $\mathcal{N}_{\mathcal{V}}(E)$ is expected to converge to the nonrandom quantity $\mathcal{N}(E)=\lim_{\mathcal{V}\to\infty}\mathcal{N}_{\mathcal{V}}(E)$ due to self-averaging~\cite{vanrossum_density_1994}, and we focus on the low-energy properties of $\mathcal{N}(E)$. The rationale of our approach lies in the connection between low-energy excitations of the Bose gas and its superfluid properties. As anticipated in the introduction, Landau's criterion for superfluidity requires the excitation dispersion relation $E(k)$ to be such that $v_c\equiv\textrm{min}_k [E(k)/(\hbar k)]>0$. In the homogeneous case, the BdGE~(\ref{eq:BdGE}) has plane-wave solutions that cross over from a quadratic free-particle-like dispersion for $k\gg1/\xi$ [\ie $E(k)\gg\mu=U$] to a linear phononlike dispersion $E(k)\sim \hbar v_s^0 k$ for $k\ll1/\xi$, with $v_s^0=\sqrt{U/m}$ identified as the sound velocity~\cite{pitaevskii_book_2003}. Hence, $v_c=v_s^0$ and the weakly interacting Bose gas ($U>0$) is superfluid. In the presence of disorder, Landau's criterion needs to be examined with care due to the broken translation invariance. For weak disorder, the spectral broadening of low-energy Bogoliubov excitations is negligible in comparison to their energy shift~\cite{gaul_bogoliubov_2011}, so that the notion of dispersion relation remains meaningful. In this regime, the linear low-energy dispersion survives with a speed of sound $v_s$ that is reduced by disorder~\cite{gaul_speed_2009}. This implies
\begin{align}\label{eq:NEsound}
\mathcal{N}(E)&=\frac{\alpha_d}{(2\pi)^d}k(E)^d\sim\frac{\alpha_d}{(h v_s)^d} E^d \quad (E\to0^+),
\end{align}
where $\alpha_d$ is the volume of the unit ball in $d$ dimensions, and $k(E)$ is the inverted dispersion relation. The CDOS thus increases in the presence of weak disorder, but $\mathcal{N}(E)E^{-d}$ remains bounded for $E\to0^+$. At the transition to the BG insulator, however, sound is expected to be suppressed in a handwaving application of Landau's criterion. Then, the assumption of a well-defined dispersion $E(k)$ satisfying $\lim_{k\to 0}[E(k)/(\hbar k)]=0$ implies that $\mathcal{N}(E)E^{-d}$ diverges at low energy. In agreement with this scenario, the density of states $\mathcal{D}(E)=d\mathcal{N}/dE$ has been shown to develop a power-law divergence in the 1D BG phase~\cite{ziman_localization_1982, fontanesi_superfluid_2009, fontanesi_mean-field_2010}. Accordingly, we use the {\it boundedness of $\mathcal{N}(E)E^{-d}$ when $E\to0^+$ as a criterion of superfluidity}, and the onset of a divergence as a signature of the SF-BG transition. We observed that for large 2D systems as considered here the computation of superfluid fractions with twisted boundary conditions~\cite{fontanesi_mean-field_2010} poses serious numerical challenges, and the CDOS results prove more reliable and accurate.


To characterize the SF-BG transition, we computed the Bogoliubov CDOS in finite but large systems of size $\mathcal{V}=L^2$ with periodic boundary conditions, and averaged over disorder configurations to obtain reliable estimates of $\disav{\mathcal{N}_\mathcal{V}}(E)$, \ie $\mathcal{N}(E)$ up to residual finite-size effects. Equations~(\ref{eq:GPE}) and (\ref{eq:BdGE}) were discretized on a square lattice with spacing $\ell=L/n_\ell$, while aiming at $t=\hbar^2/(2m\ell^2)\gg U, \Delta, E_c$ in order to emulate the continuum limit. Unless stated otherwise, we simulated systems of size $512\eta \times 512 \eta$, and  we chose $t=4E_c$ (\ie $\eta=2\ell$) to resolve energies $E\ll U,\Delta,E_c$ despite finite-size cutoffs.

\begin{figure}[t]
\includegraphics[width=7.8cm]{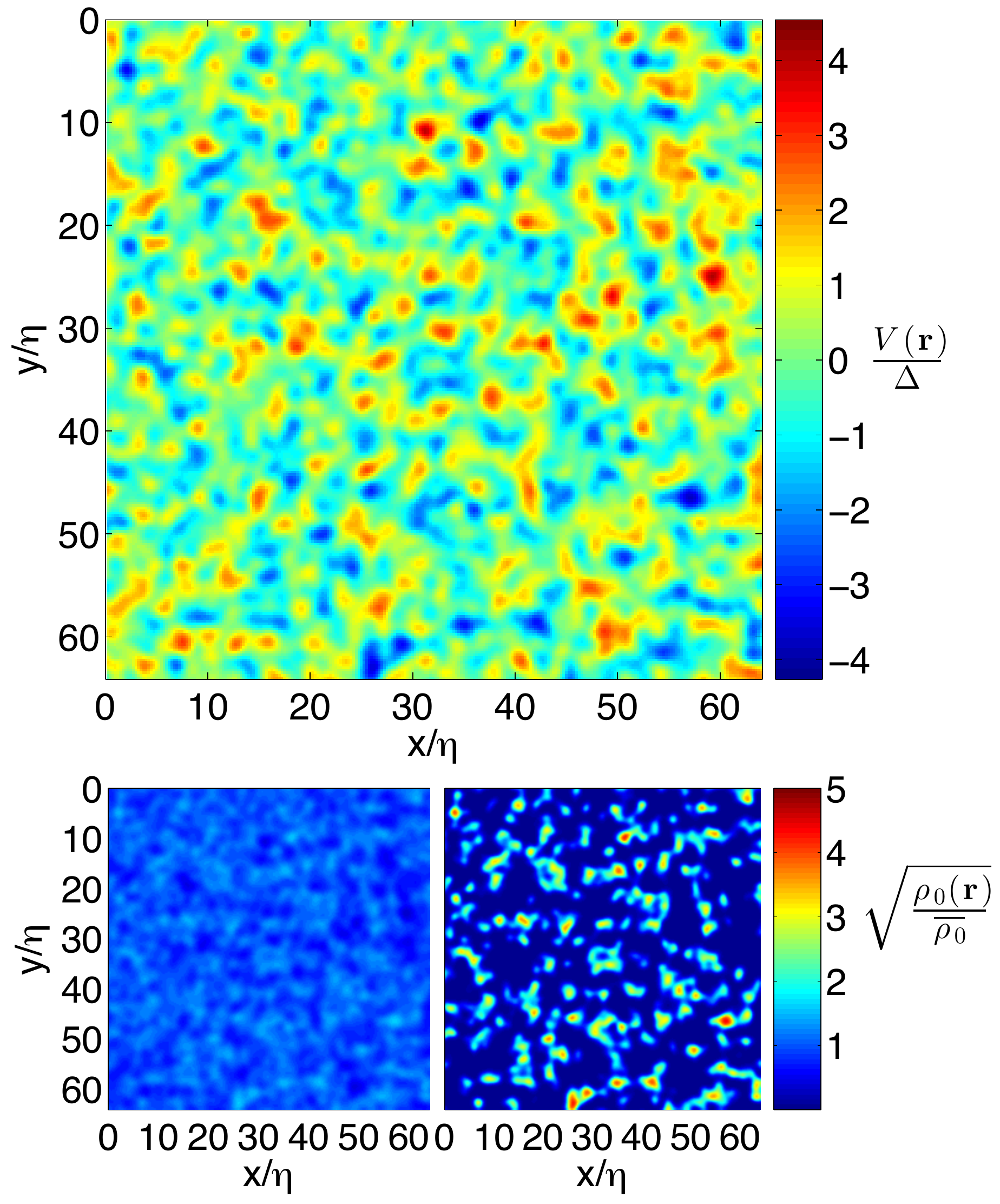}
\caption{\label{fig:potential_and_density}(Color online) Upper panel: disorder configuration with correlation length $\eta=4\ell$. Lower panels: corresponding ground-state density profiles $\rho_0(\vecr)$ for an interaction strength of $U=g\disav{\rho_0}=1.6E_c$ and disorder amplitudes $\Delta=0.8 E_c$ (left) and $\Delta=9.6E_c$ (right).}
\end{figure}

For each disorder configuration, the density $\rho_0(\vecr)$ was calculated with a conjugate-gradient technique~\cite{modugno_bose-einstein_2003}. Figure~\ref{fig:potential_and_density} shows a disorder configuration and the resulting density for two disorder amplitudes. Subsequently we computed the \textit{local} CDOS $\mathcal{N}_\mathcal{V}(\vecr,E)=\sum_{0<E_j\leq E}p_j(\vecr)$, where $p_j(\vecr)=u_j^2(\vecr)-v_j^2(\vecr)$, for a set of positions $\vecr$, using a kernel polynomial method (KPM)~\cite{weise_kernel_2006, saliba_superfluid-insulator_2013}. The latter offers an efficient alternative to the (partial) diagonalization of $\Lgp$, \eg via Lanczos-type techniques, which becomes prohibitive for the system sizes under scrutiny.  The particular weighting $p_j(\vecr)$ of the (real-valued) Bogoliubov modes stems from their biorthogonality relations~\cite{castin_low-temperature_1998, mora_extension_2003, saliba_superfluid-insulator_2013} and corresponds to a trace over particle-hole ($u$-$v$) space. This weighting also implies that the CDOS for a configuration may be obtained as the spatial average $\mathcal{N}_\mathcal{V}(E)=\mathcal{V}^{-1}\int d\vecr \mathcal{N}_\mathcal{V}(\vecr,E)$.

\begin{figure}[t!]
\includegraphics[width=8.0cm]{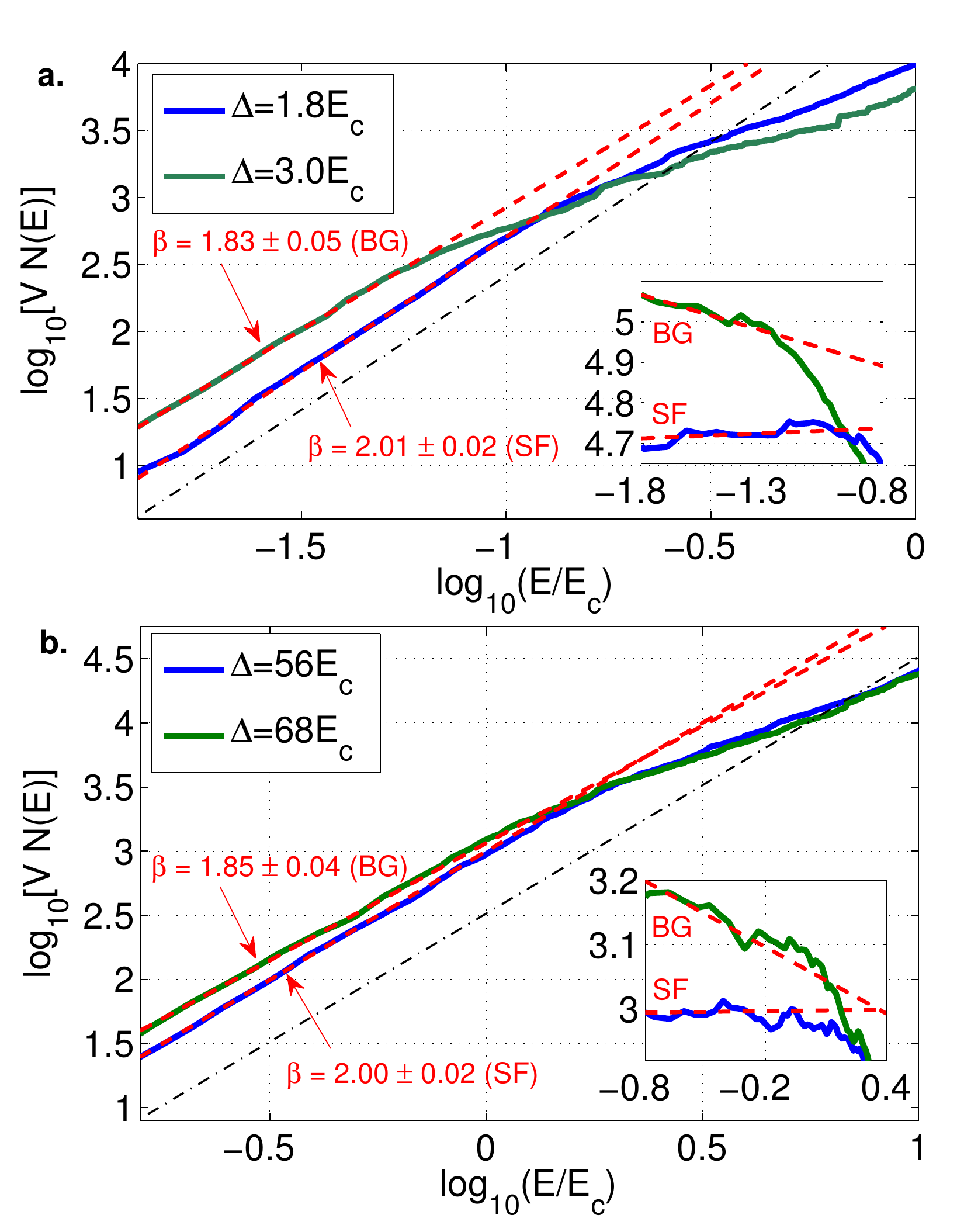}
  \caption{(Color online) Averaged CDOS $\disav{\mathcal{N}_\mathcal{V}}(E)\simeq \mathcal{N}(E)$, multiplied by the volume $\mathcal{V}$, for (a) $U=0.4E_c$ and (b) $U=32E_c$. The red dashed lines are linear fits to the low-energy part of the loglog data. The blue and green data bracket the critical disorder strength, as revealed by a change in the $\beta$ exponent of the power-law asymptotics (see Fig.~\ref{fig:Exponent}). The black dash-dotted lines show the $\Delta=0$ asymptotics $\mathcal{V}\mathcal{N}(E)\sim\mathcal{V}\alpha_2(h v_s^0)^{-2} E^2$ [see Eq.~(\ref{eq:NEsound})]. Insets: $\log_{10}[\mathcal{V}E_c^2\disav{\mathcal{N}_\mathcal{V}}(E)E^{-2}]$ \textit{vs.} $\log_{10}(E/E_c)$. The dashed lines reflect the fits of the main panels, and negative slopes mark a divergence of $\mathcal{N}(E)E^{-2}$ as $E\to0^+$, \ie an insulating behavior.}
\label{fig:Comparative_CDOS}
\end{figure}

In our KPM scheme, the local CDOS is expanded as
\begin{align}\label{eq:ChebyshevExpansionLCDOS}
\mathcal{N}_\mathcal{V}(\vecr,E)
&=\frac{\mu_0(\vecr)}{\pi}\arcsin\left(\frac{E}{\Emax}\right)\nonumber\\
&\quad - \sum_{p=1}^{+\infty}\frac{\mu_{2p}(\vecr)}{p\pi}\sin\left[2p \arccos\left(\frac{E}{\Emax}\right)\right]\nonumber\\
&\quad -2\phi_a(\vecr)\phi_0(\vecr),
\end{align}
where the $\mu_n(\vecr)$ are Chebyshev moments~\cite{weise_kernel_2006} of the local Bogoliubov DOS, given by
\begin{align}\label{eq:mun}
\mu_n(\vecr)&=\begin{pmatrix}\bra{\vecr},\bra{\vecr}\end{pmatrix}T_n\left(\Lgp/\Emax\right)\begin{pmatrix}\ket{\vecr}\\\ket{\vecr}\end{pmatrix}.
\end{align}
Above, the $T_n$ are Chebyshev polynomials of the first kind, and $\Emax$ is a scaling factor slighty larger than the maximum eigenvalue of the discretized $\Lgp$, which we calculated by power iteration for each configuration. Equation~(\ref{eq:mun}) allows an iterative computation of $\mu_n(\vecr)$, $n=0,1,\dots$, that requires only one sparse-matrix vector multiplication for each new order $n$. We truncated expansion (\ref{eq:ChebyshevExpansionLCDOS}) at order $n_\mathrm{max}=10^6$ and used a Jackson kernel to damp Gibbs oscillations~\cite{weise_kernel_2006}, thereby achieving a level broadening well below $10^{-3}E_c$ even for the largest disorder amplitudes, for which $\Emax$ reached $400 E_c$. Expression~(\ref{eq:mun}) entails a contribution from the zero eigenspace of $\Lgp$~\cite{castin_low-temperature_1998, mora_extension_2003}, which is canceled by the last line of Eq.~(\ref{eq:ChebyshevExpansionLCDOS}), where $\phi_0(\vecr)=\sqrt{\rho_0(\vecr)/N_0}$, with $N_0$ the number of bosons, and $\phi_a(\vecr)$ is an anomalous term, which can be calculated explicitly~\cite{castin_low-temperature_1998, mora_extension_2003}. We found the $2\phi_a(\vecr)\phi_0(\vecr)$ term to have a negligible impact outside the low-energy range where only very few ($\lesssim$\,10) states are accumulated in $\mathcal{N}_\mathcal{V}(\vecr,E)$, and we omitted the counterterm in Eq.~(\ref{eq:ChebyshevExpansionLCDOS}) accordingly. Finally, rather than averaging $\mathcal{N}_\mathcal{V}(\vecr,E)$ over $\vecr$ for each configuration, we directly performed the disorder average at position $\vecr=\bvec{0}$ to obtain $\disav{\mathcal{N}_\mathcal{V}}(E)$.

\begin{figure}
  \includegraphics[width=8.7cm]{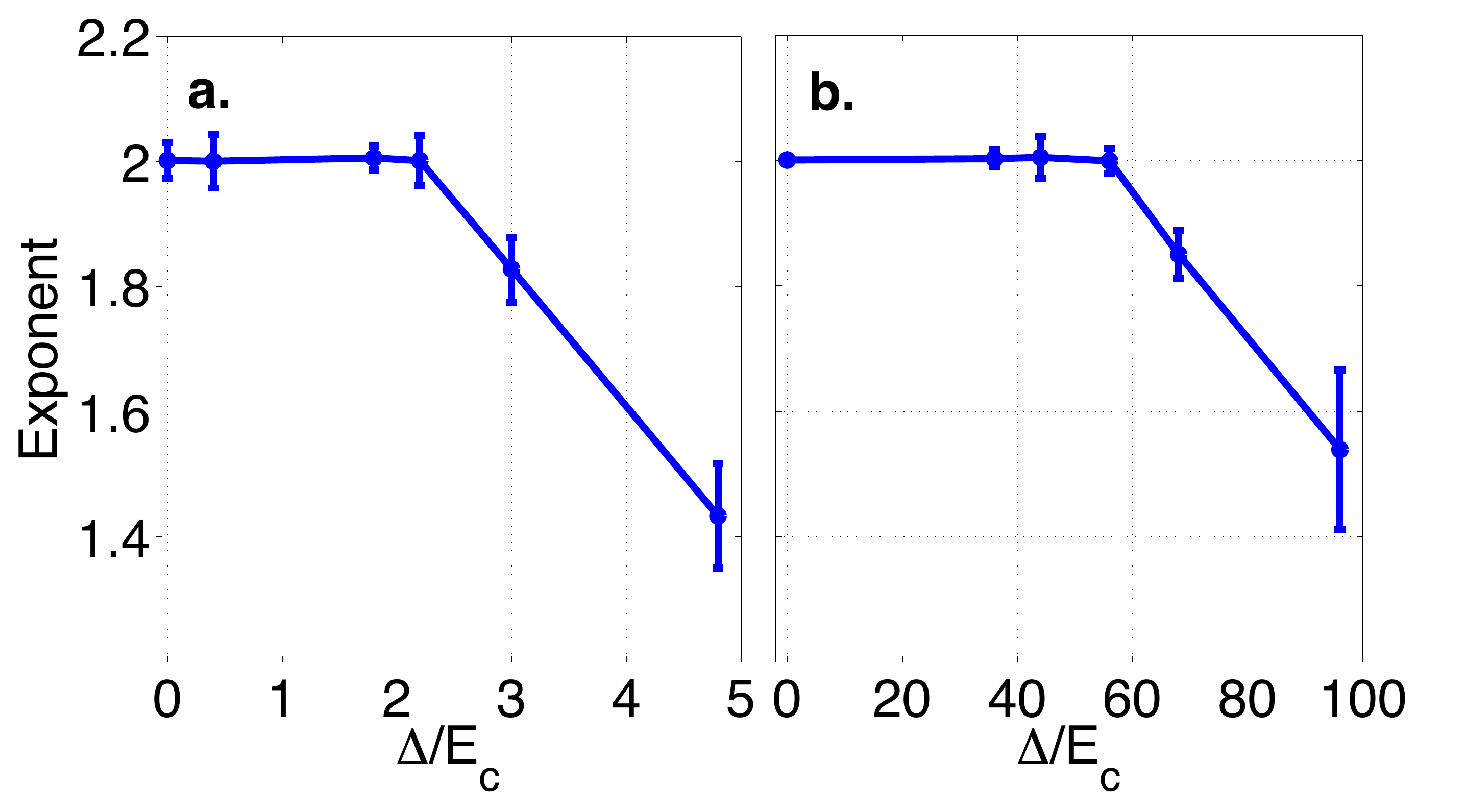}
  \caption{(Color online) Power-law exponent $\beta$ versus disorder strength $\Delta$, as obtained from fits to the low-energy CDOS data (see Fig.~\ref{fig:Comparative_CDOS}), for (a) $U=0.4E_c$ and (b) $U=32 E_c$. The error bars are 95\% confidence intervals of the fits.}
  \label{fig:Exponent}
\end{figure}

With the above procedure, we computed the Bogoliubov CDOS for various interaction strengths $U$ and for increasing disorder strength $\Delta$, averaging over 200 to 1000 configurations for each $(U,\Delta)$ pair. For all values of $U$, we found (i) a gapless spectrum at all disorder strengths (down to finite-size gaps of about $10^{-2}E_c$ to $10^{-1}E_c$ by increasing values of $U$), as expected for the SF and BG phases; (ii) a quadratic and enhanced CDOS $\mathcal{N}(E)$ at low energies, reflecting a speed of sound reduced by disorder; and (iii) at a critical disorder strength, the onset of a power-law behavior $\mathcal{N}(E)\sim E^\beta$ with (disorder-dependent) exponent $\beta<2$, signaling the loss of sound. Figures~\ref{fig:Comparative_CDOS}(a) and~\ref{fig:Comparative_CDOS}(b) show the disorder-averaged CDOS obtained for a set of representative $(U,\Delta)$ pairs in the white-noise (WN, $U\ll E_c$) and Thomas-Fermi (TF, $U\gg E_c$) regimes. The red dashed curves are linear fits to the log-log data at low $E$. Figure~\ref{fig:Exponent} displays the corresponding slopes $\beta$, and demonstrates a threshold behavior that we associate with the SF-BG transition.

\begin{figure}
\includegraphics[width=8cm]{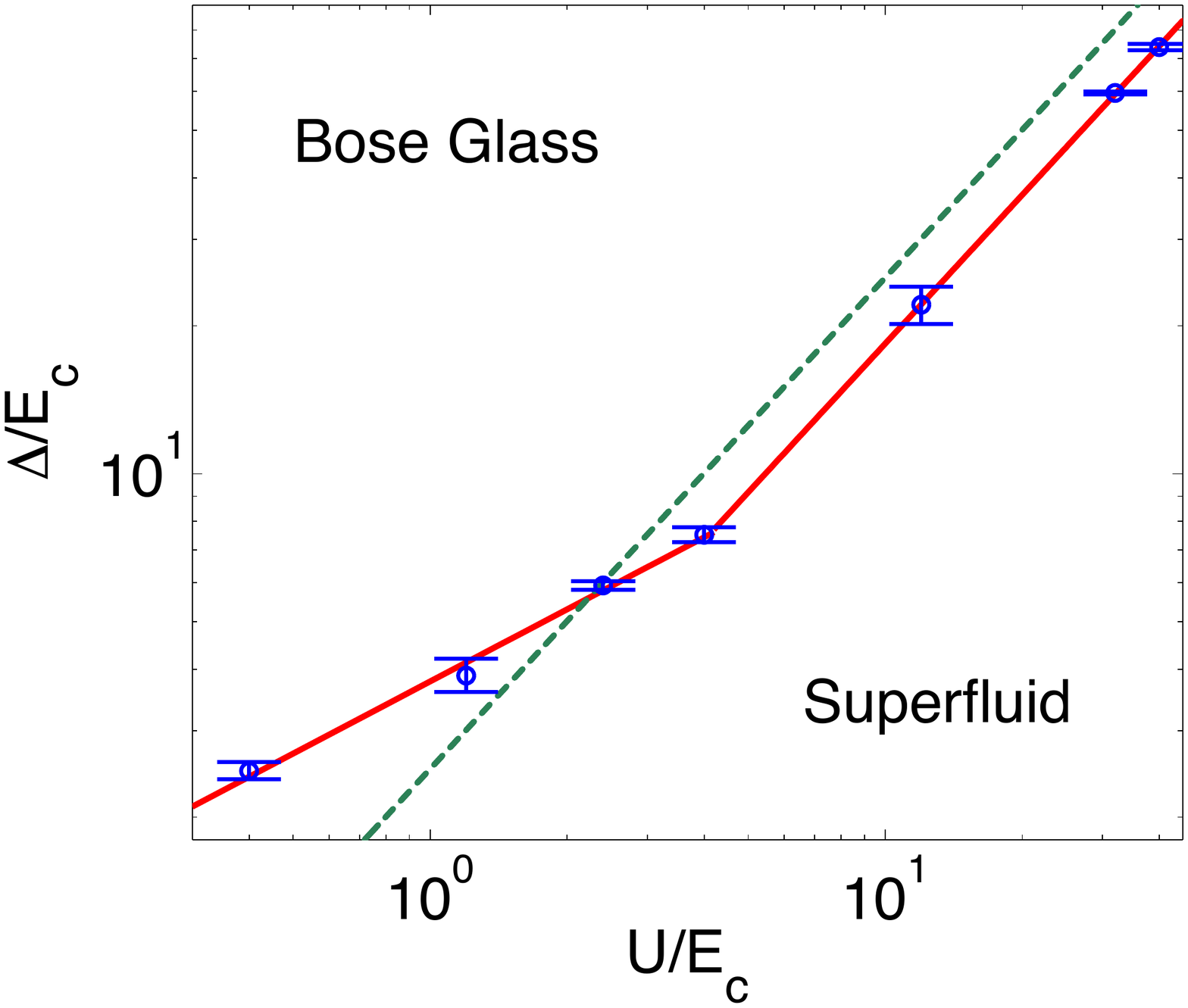}
\caption[format=normal]{(Color online) Phase diagram of the 2D Bose gas as a function of interaction and disorder. The green dashed line shows the percolation threshold for the density $\rho_0(\vecr)$ based on the Thomas-Fermi limit of Eq.~(\ref{eq:GPE}). The red line represents linear fits to the four leftmost and three rightmost data points on log-log scale.}
\label{fig:Phase_Diag}
\end{figure}

To determine the critical disorder for each $U$, the $\beta$ data was interpolated linearly and a threshold $\beta$ value set at 1.96 to account for the typical error bars of $2\%$ found close to the transition (see Fig.~\ref{fig:Exponent}). The resulting phase diagram is shown in Fig.~\ref{fig:Phase_Diag}. The error bars on the data reflect the uncertainty on the critical $\Delta$ inherited from the $\beta$ exponents. In the WN regime the boundary follows a power law $\Delta/E_c = \zeta \left( U/E_c\right)^\alpha$, with $\alpha = 0.49 \pm 0.13$ and $\zeta=3.78 \pm 0.45$, in agreement with the square-root dependence expected at the mean-field level, irrespective of Gaussian statistics~\cite{falco_weakly_2009, fontanesi_mean-field_2010}. In the TF regime we find a power $\alpha=1.01 \pm 0.10$ that is also consistent with the linear behavior expected from mean-field scaling arguments~\cite{falco_weakly_2009, fontanesi_mean-field_2010}. The critical ratio $\Delta/U=1.83\pm0.37$ in this regime lies below the TF percolation threshold $\Delta/U=\sqrt{2\pi}$ of $\rho_0(\vecr)$ in Gaussian 2D disorder~\cite{isichenko_percolation_1992}. In other words, the classical percolation of the ground-state density is not sufficient to ensure superfluidity. These findings agree with those of a recent Monte-Carlo study of the $T>0$ phase diagram in speckle disorder~\cite{carleo_universal_2013}, and suggest that the notion of superfluid puddles in a percolation picture of the transition~\cite{iyer_mott_2012, krinner_superfluidity_2013, niederle_superfluid_2013} should be characterized with care. It is also worth comparing our results to the $T=0$ study of Ref.~\cite{astrakharchik_phase_2013}. While the superfluid fractions found therein for the WN regime are consistent with a square-root law, no data were presented for the TF regime where a classical percolation analysis applies. Moreover, the data of Ref.~\cite{astrakharchik_phase_2013} were obtained for systems of a few tens of correlation lengths and extrapolated with an \textit{ad hoc} scaling law. By contrast, the system sizes achieved here and the threshold behavior observed in the power-law CDOS allow us to locate the phase boundary with an accuracy of $10\%-20\%$ with the presently available $\Delta$ values. It is worth noting that a reduction of the linear system size $L$ lifts the low-energy finite-size cutoff by the same factor, due to the linear dispersion at the transition. We found that in systems of $128\eta\times128\eta$ the asymptotics of Fig.~\ref{fig:Comparative_CDOS} were barely emerging, which typically lead to an underestimation of the critical $\Delta$. Quite generally, a finite size is expected to limit the resolution on the critical $\Delta$ by masking the departure from $\beta=2$ at arbitrarily low energy.

In conclusion, we analyzed the cumulative density of states $\mathcal{N}(E)$ of the Bogoliubov excitations of disordered bosons in 2D. We found power-law asymptotics $\mathcal{N}(E)\sim E^\beta$ at low energy for all disorder strengths, with a sharp threshold behavior in the exponent $\beta$ indicating a transition to the Bose-glass phase. Our numerical results provide a quantitative picture of the $T=0$ mean-field phase diagram in the white-noise and Thomas-Fermi regimes, which should be valuable for both the analysis of present experiments~\cite{deissler_correlation_2011, allard_effect_2012, krinner_superfluidity_2013} and the identification of other signatures of the 2D phase transition, \eg in coherence~\cite{saliba_superfluid-insulator_2013} or localization properties~\cite{gurarie_excitations_2008, zuniga_bose-glass_2013}.

We thank N. Prokof'ev, G. Carleo and G. Bo\'eris for fruitful discussions. This work was supported by the Swiss National Science Foundation through Project No.~200020\_149537.

\bibliography{DOSpaper_20140815submission}

\end{document}